\documentclass[showpacs,preprintnumbers,preprint]{revtex4}
\usepackage{graphicx,psfrag}
\usepackage{bm}

\begin{document}

\preprint{APS/000}

\title{Chaotic beats  in a nonautonomous system governing
second-harmonic generation of light}

\author{I. \'{S}liwa}
 \email{izasliwa@amu.edu.pl}
\author{P. Szlachetka}
 \email{przems@amu.edu.pl}
\author{K. Grygiel}
 \email{grygielk@amu.edu.pl}
\affiliation{Nonlinear Optics Division, Institute of Physics,
A. Mickiewicz University,
  Umultowska 85, PL 61-614 Pozna\'{n}, Poland }

\date{\today}

\begin{abstract}
The letter proposes a procedure for generation and control chaotic beats in a dynamical system
being initially in the periodic state.  The dynamical system describes a simple
nonlinear optical process -- second-harmonic generation of light. The periodic states
of the system have been found in analytical forms. We also investigate some aspects of
synchronization of chaotic beats in  two systems,  detuned in the pump fields.

\end{abstract}


\maketitle
\begin{figure}
\includegraphics[width=10cm,height=4cm,angle=0]{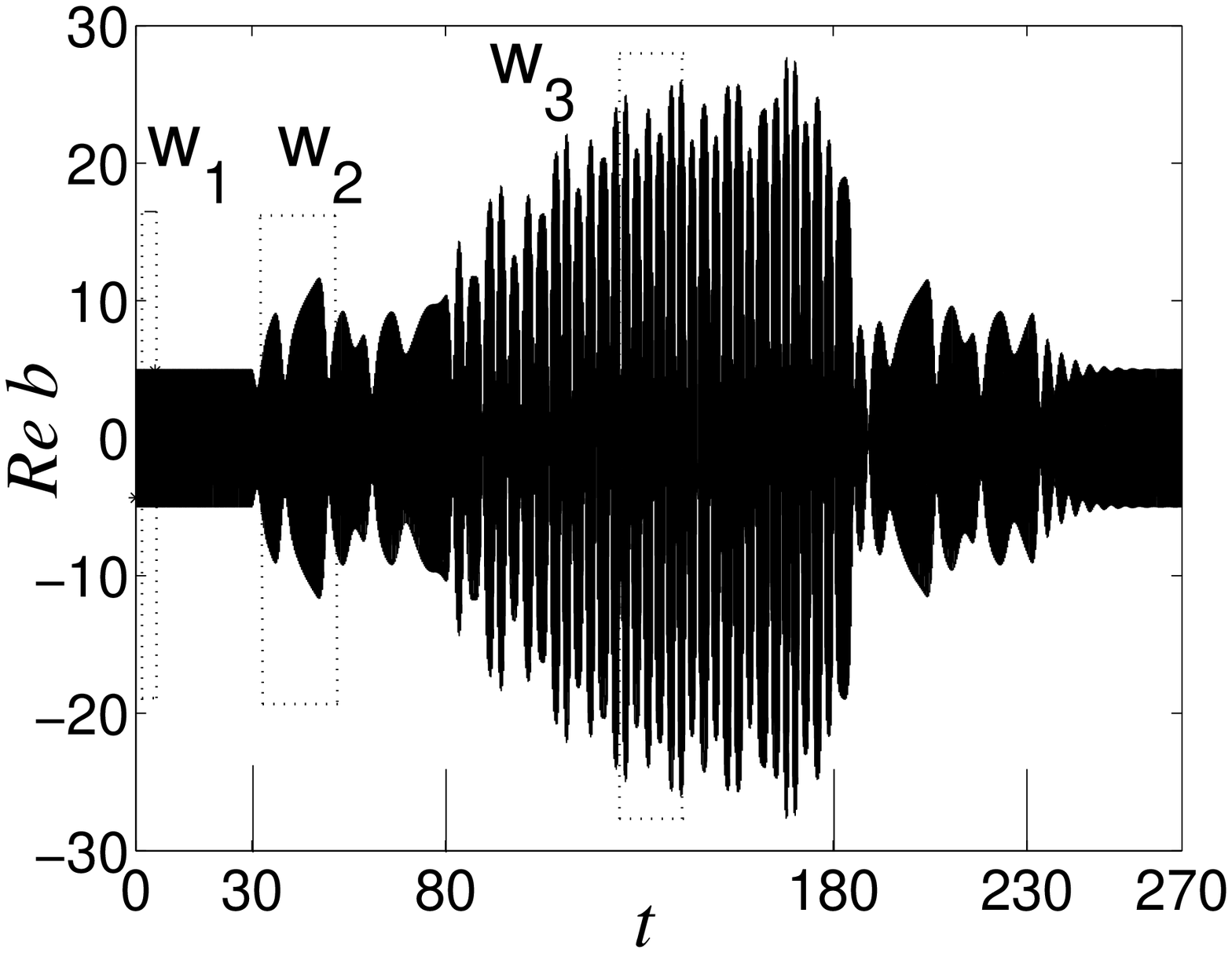}
\includegraphics[width=10cm,height=4cm,angle=0]{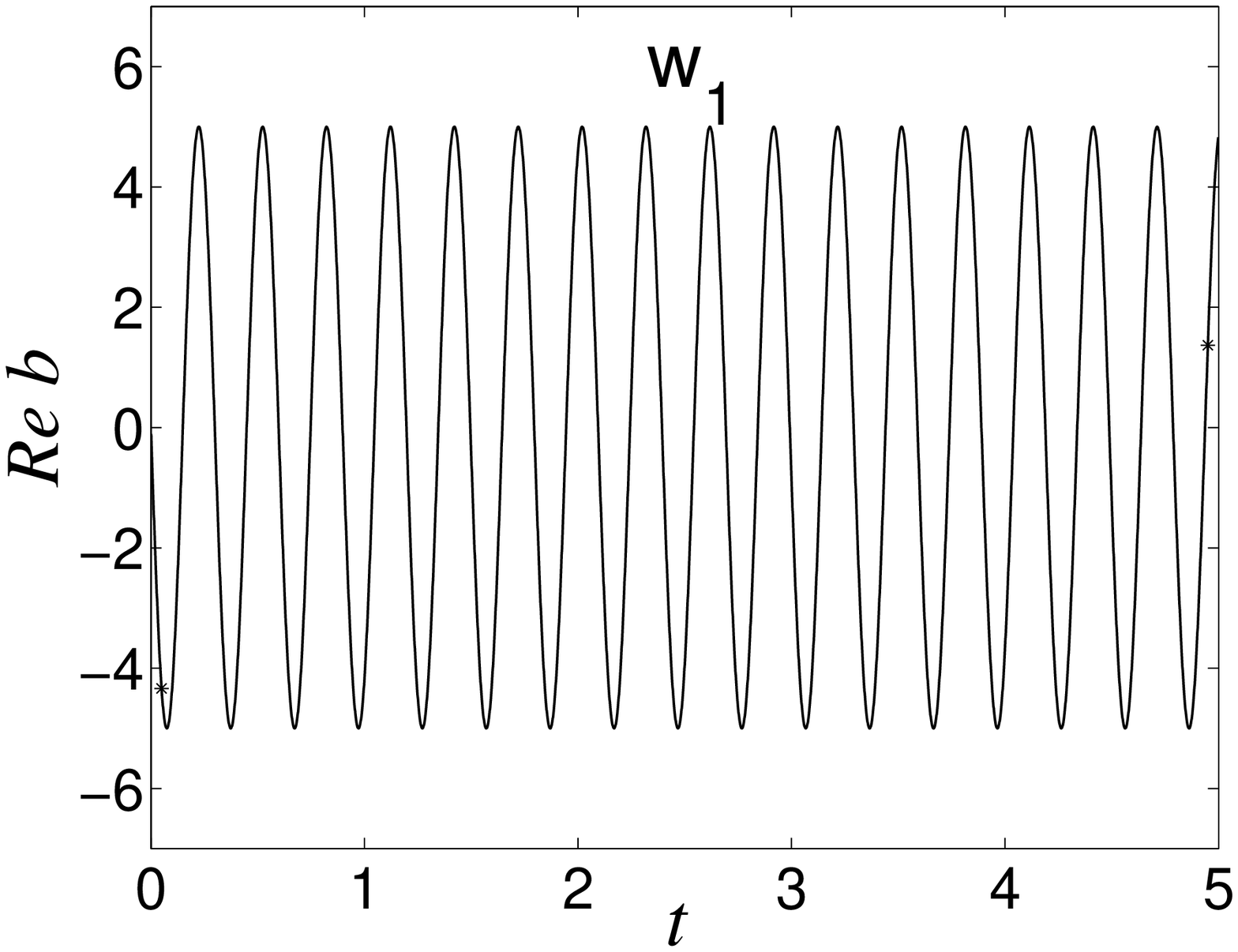}
\includegraphics[width=10cm,height=4cm,angle=0]{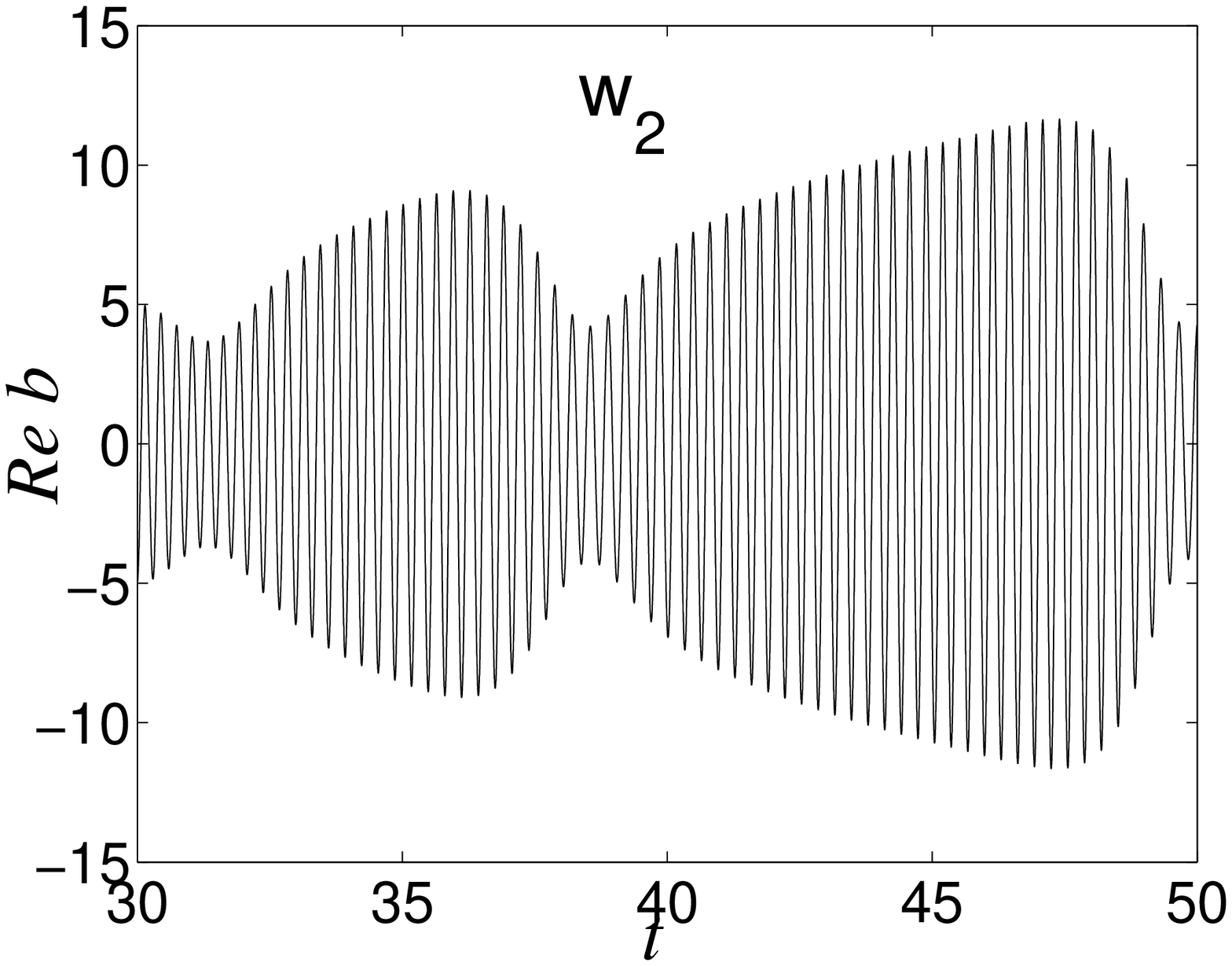}
\includegraphics[width=10cm,height=4cm,angle=0]{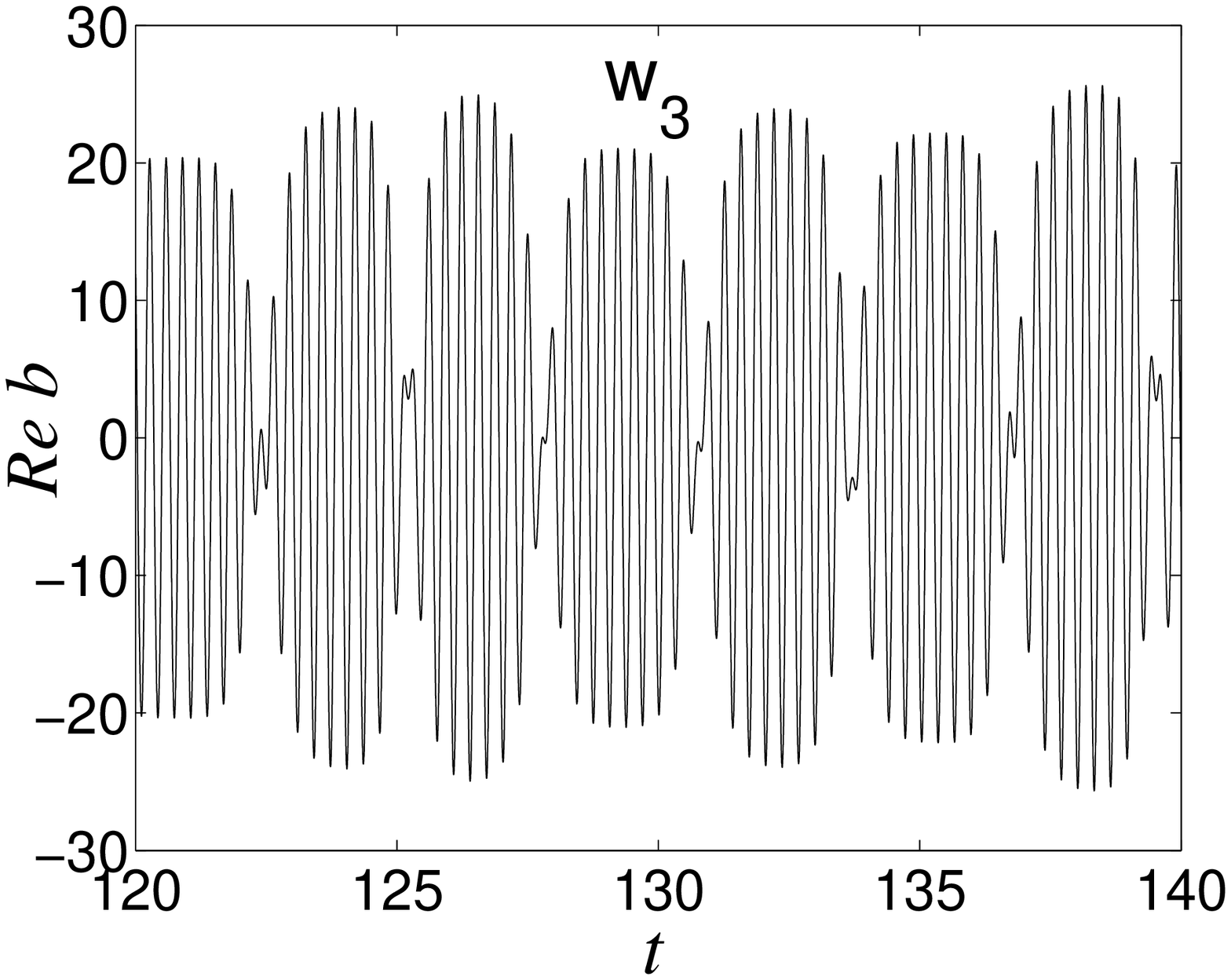}
\caption{}
 \label{fig.1}
\end{figure}

It is known that a nonlinear system can produce vibrations having the structure of intricate
revivals and collapses [Minorsky,1962; Eberly, 1980].
If the system is chaotic, this type of vibrations are referred to as
chaotic beats [Grygiel \& Szlachetka, 2002].  In the simplest cases chaotic beats can
 be interpreted as a "signal"
with chaotic envelopes and a stable fundamental frequency. In much
more complex systems not only the envelopes but also frequencies are
chaotically modulated. Recently, chaotic beats have been
theoretically and experimentally studied in Chua's circuits
[Cafanga \& Grassi, 2004,2005]. In this paper we show  how to generate
chaotic beats in a dynamical system being initially in a
periodic state. To generate the  beats we use the following
dynamical system in the complex variables system $a$ and $b$
(four equations in real variables)[Mandel \& Erneux , 1982;
Pe\v{r}ina, 1991]:
\begin{eqnarray}
\label{e1}
\frac{da}{dt}&=&-i\omega a -\gamma a +\epsilon a^{*}b +F e^{-i\Omega t}\,,\\
\label{e2}
\frac{db}{dt}&=&-i2\omega b  -\frac{1}{2}\epsilon a^{2}\,.
\end{eqnarray}
Physically, the equations describe second harmonic-generation of light in the so-called
good frequency conversion limit.
 The complex variables $a$ and $b$ are the amplitudes of the fundamental
 and second-harmonics modes, respectively.   The interaction between the modes
  takes place via a nonlinear crystal placed within a
 Fabry-Pe\'rot interferometr. The quantities $\omega$ and $2\omega$  are the
  frequencies of the fundamental and second-harmonic modes, respectively.
 The nonlinear coupling coefficient $\epsilon$
  is proportional to the second-order nonlinear susceptibility. The parameter
 $\gamma$  is a damping constant. Moreover, the system is pumped by an external field
$F e^{-i\Omega t}$, where $F$  is an electric
 field amplitude at the frequency $\Omega$.
Henceforth, all the parameters, that is $\omega$, $\epsilon$, $F$, and  $\Omega$
 are taken to be real.
\\
It is easy to find that for the fixed parameters $\omega$,
$\gamma$, $\epsilon$ and $F$, the system (\ref{e1})-(\ref{e2})
has two pairs of periodic solutions provided that
$\Omega_{\pm}=\omega\pm (\epsilon F/2\gamma)$. The first pair has
the form
\begin{eqnarray}
\label{e3}
a_{+}(t)&=&\frac{F}{\gamma} e^{-i\Omega_{+} t}\,,\\
\label{e4}
b_{+}(t)&=&-\frac{i}{2}\frac{F}{\gamma}e^{-i2\Omega_{+} t}\,.
\label{e5}
\end{eqnarray}
The second pair is given by
\begin{eqnarray}
\label{e6}
a_{-}(t)&=&\frac{F}{\gamma} e^{-i\Omega_{-} t}\,,\\
\label{e7}
b_{-}(t)&=&\frac{i}{2}\frac{F}{\gamma}e^{-i2\Omega_{-} t}\,.
\label{e8}
\end{eqnarray}
It is easy to note that in the phase plane $(Re\, a(t),Im\, a(t))$ the periodic solutions
$a_{+}(t)$ and $a_{-}(t)$
 satisfy the same phase equation (circle)
\begin{eqnarray}
\label{phase1}
[Re\, a(t)]^{2}+[Im\, a(t)]^{2}=\frac{F^{2}}{\gamma^{2}}\,.
\end{eqnarray}
The differences are only in the angular velocities $\Omega_{\pm}$ - one circle is
drawn faster than the other.
 A similar behavior is observed in the phase plane $( Re\, b(t),Im\, b(t))$ for
 the functions $b_{+}(t)$ and  $b_{-}(t)$.
 Here, the phase equation has the form
 \begin{eqnarray}
\label{phase2}
[Re\, b(t)]^{2}+[Im\, b(t)]^{2}=\frac{F^{2}}{4\gamma^{2}}\,.
\end{eqnarray}
Purely formally, the system (\ref{e1})-(\ref{e2}) has the periodic solutions
(\ref{e3})-(\ref{e8}) if $\Omega=\omega\pm (\epsilon F/2\gamma)$ and if it starts
 from the initial conditions $a(0)=F/\gamma$ and $b(0)=\mp iF/2\gamma$.\\
The phase curves (\ref{phase1}) and (\ref{phase2}) represent a steady state of the system
(\ref{e1})-(\ref{e2}) or its unstable periodic orbit. The type of behavior
depends on the parameters of the system. By way of example, for $\omega=10$, $\gamma=0.5$,
$\epsilon=0.1$,  $F=5$ and $\Omega=10.5$ (or $\Omega=9.5$) the system in the phase planes
tends to make circles $[Re\, a(t)]^{2}+[Im\, a(t)]^{2}=10^{2}$  and
$[Re\, b(t)]^{2}+[Im\, b(t)]^{2}=5^{2}$ irrespectively of the values of the
initial conditions $a(0)$
and $b(0)$. In particular, if the system starts from the initial conditions $a(0)=10$
and $b(0)=\mp 5i$ the phase points draw simply the circles (attractors)
with the radii $10$ and $5$.\\
If, in the above example, we put $\gamma=0.1$ instead of $\gamma=0.5$ the new circles
 $[Re\, a(t)]^{2}+[Im\, a(t)]^{2}=100^{2}$ and $[Re\, b(t)]^{2}+[Im\, b(t)]^{2}=50^{2}$ do not posses the
attractor's properties. The system moves on the circles only if  it starts
from the initial conditions $a(0)=100$ and $b(0)=\mp 50i$. Otherwise,
 the system tends to make other circles of unknown analytical forms.\\
The existence of periodic solutions suggests that in their neighbourhoods
the revivals and collapses (beats) appear if the periodicity of the system is suitably disturbed. The
transition from the periodic states  (\ref{e3})-(\ref{e7}) to beats (quasiperiodic state)
 is accomplished, if one introduces the detuning of frequencies
 $\Omega_{\pm}\neq\omega\pm (\epsilon F/2\gamma)$.
 For example, if in the system (\ref{e1})-(\ref{e2})
 $\omega=10$, $\gamma=0.5$, $\epsilon=0.1$,  $F=5$ and
   $\Omega=10.5$ ($\Omega=9.5$) the beats  occur in the dynamical variables $a(t)$ and $b(t)$
in the range $9.82<\Omega<10.25$, that is between
 the periodic states $[a(t)=10\exp(-i9.5t),  b(t)=i5\exp(-i2\cdot9.5t)]$ and
 $[a(t)=10\exp(-i10.5t), b(t)=-i5\exp(-i2\cdot10.5t)]$.
The beats created, in this way, have a typical quasiperiodic structure.
To change their nature into distinctly chaotic it is enough to suitably
decrease the damping of the system (to indicate the degree  of chaos within the beats
we have used the maximal Lyapunov exponent $\lambda$).\\
 An illustration of the transition from  a periodic state to chaotic beats and their return
 to the initial periodic state is presented in Fig.\ref{fig.1}.  As seen,   the system is periodic
($\omega=10$, $\gamma=0.5$, $\epsilon=0.1$,  $F=5$ and
   $\Omega=10.5$)   since $0\leq t<30$.
 At the time $t=30$  we change  the external frequency from
 $\Omega=10.5$ to $\Omega=9.9$
 and the system begins to generate quasiperiodic beats ($\lambda=0$).
 To get chaotic beats we decrease
 (at the time $t=80$)  the value
 of the damping constant from $\gamma=0.5$ to $\gamma=0.04$ and the
 system begins  generation of chaotic
 beats $(\lambda = 0.0184)$. If we now want to return to the  initial periodic state
  we increase the damping constant  to the value $\gamma=0.5$. This was made at the time
  $t=180$. Consequently, the system returned  to the quasiperiodic beats. And finally,
  on changing (at the time $t=230$) the frequency  from $\Omega=9.9$ into $\Omega=10.5$  the system
  returned to the originate periodic state. The durations of the individual types of vibrations
  have been  arbitrarily chosen.

\begin{figure}
\includegraphics[width=12cm,height=12cm,angle=0]{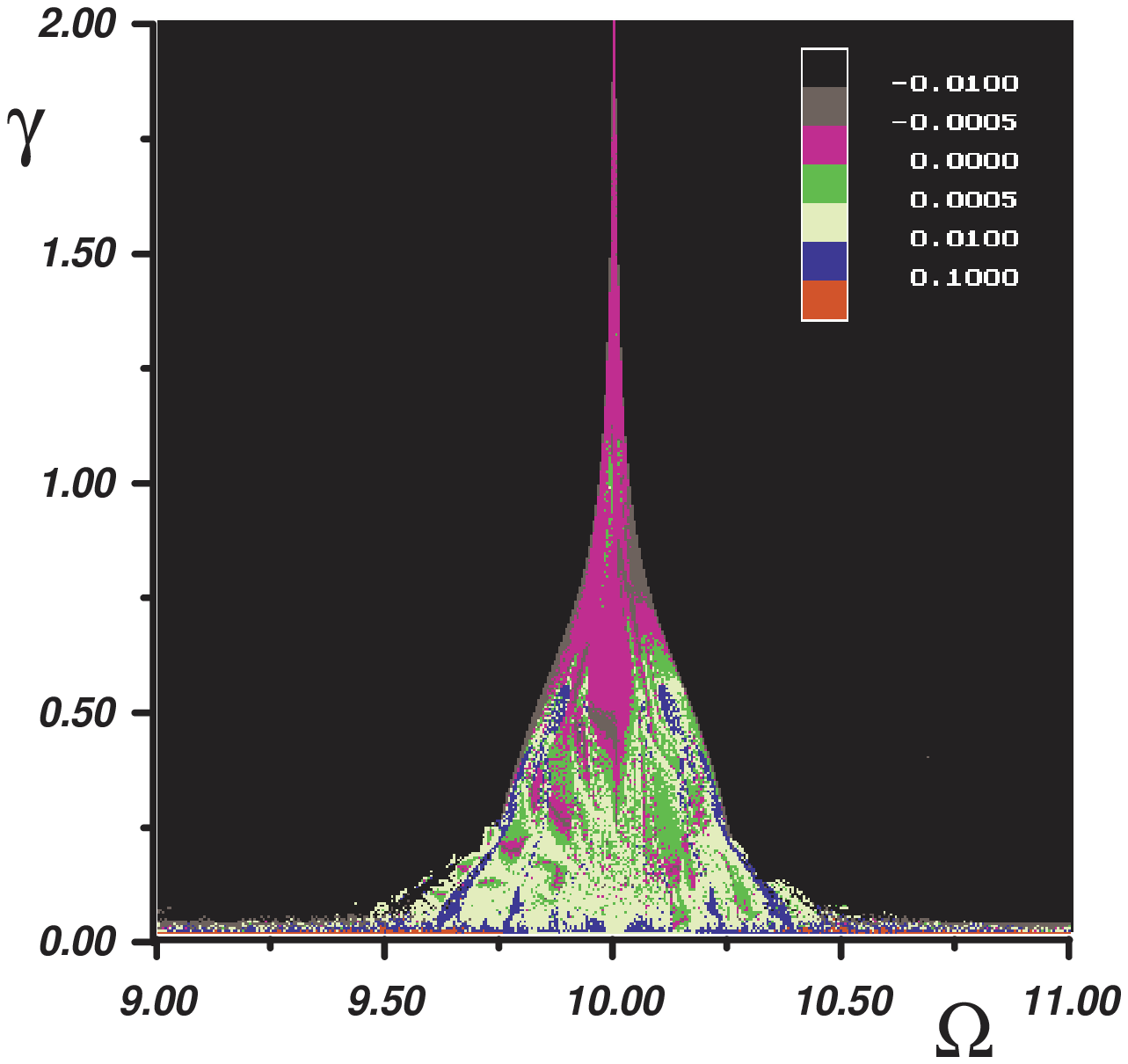}
\caption{}
 \label{fig.2}
\end{figure}

 The regions of the chaotic beats for the system (\ref{e1})--(\ref{e2}), where $\omega=10$,
 $\epsilon=0.1$,$F=5$, $0<\gamma<2$ and $9<\Omega<11$, are shown
 precisely in the Lyapunov map in the parametric space
 $(\gamma,\Omega)$ (Fig.\ref{fig.2}). The values of the maximal
 Lyapunov exponents for individual
 values of the parameters $\gamma$ and $\Omega$ are marked in an
 appropriate colour.
 The exponents have been calculated using the Wolf procedure [Wolf {\em
 et al.},1985].
  Generally, we observe chaotic beats for weak chaos -
 green and yellow color in the bottom part of the cone (Fig. 2.) When the
 damping constant
 is increased, the range of detunig
 between the frequencies $\Omega$ and $\omega$, at which
  the chaotic beats
 are generated, is diminished. Consequently, the upper part of the cones
 confines quasiperiodic beats
 ($\lambda=0$,purple color). The explanation of this fact is
 simple - a growing damping stabilizes the system by delimitation of
 the region
 of chaos strictly to the near resonance case $(\Omega \approx \omega)$. The space outside the cone
 corresponds to
 periodic states.

\begin{figure}
\includegraphics[angle=90,width=8cm,height=12cm,angle=0]{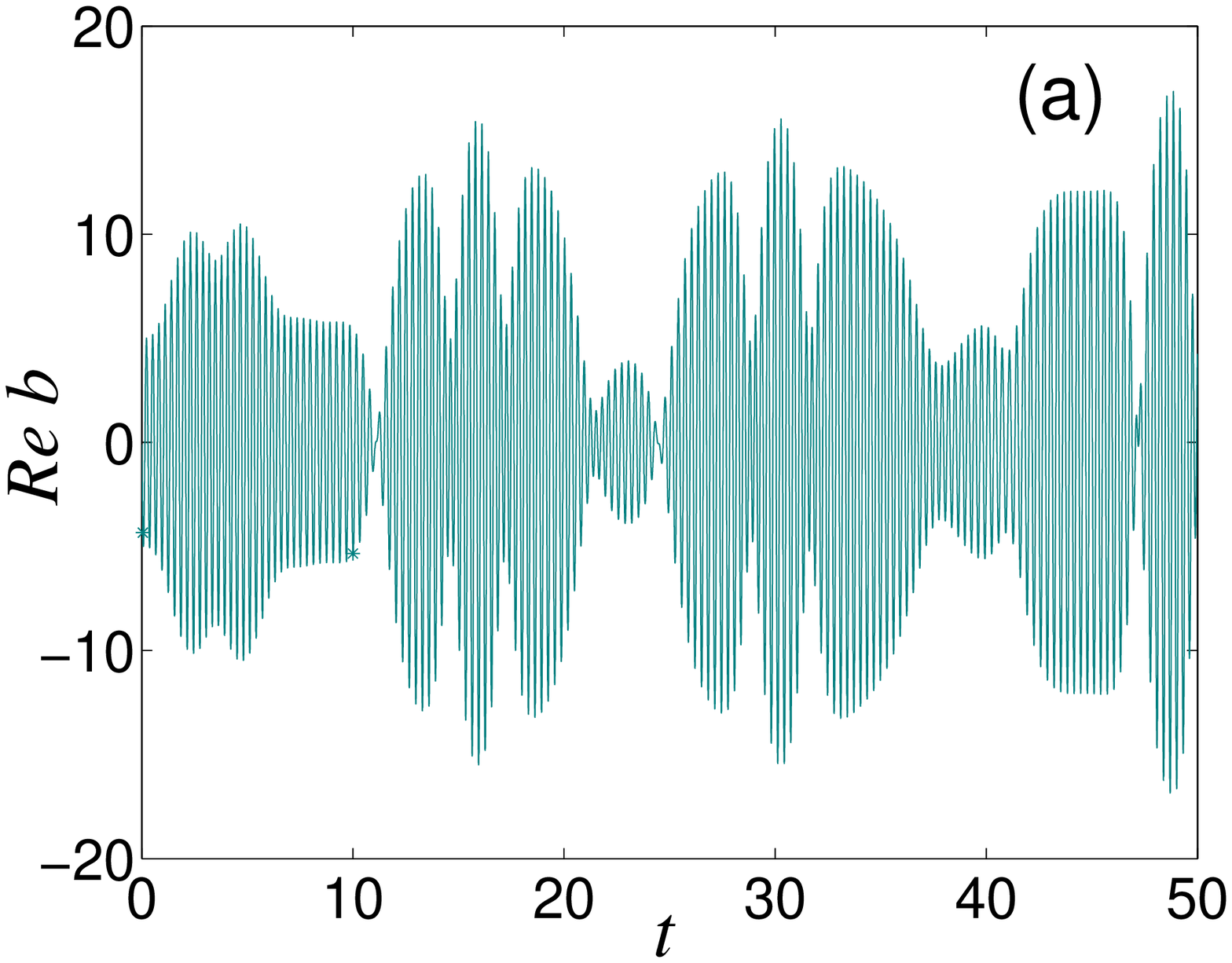}
\includegraphics[angle=90,width=8cm,height=12cm,angle=0]{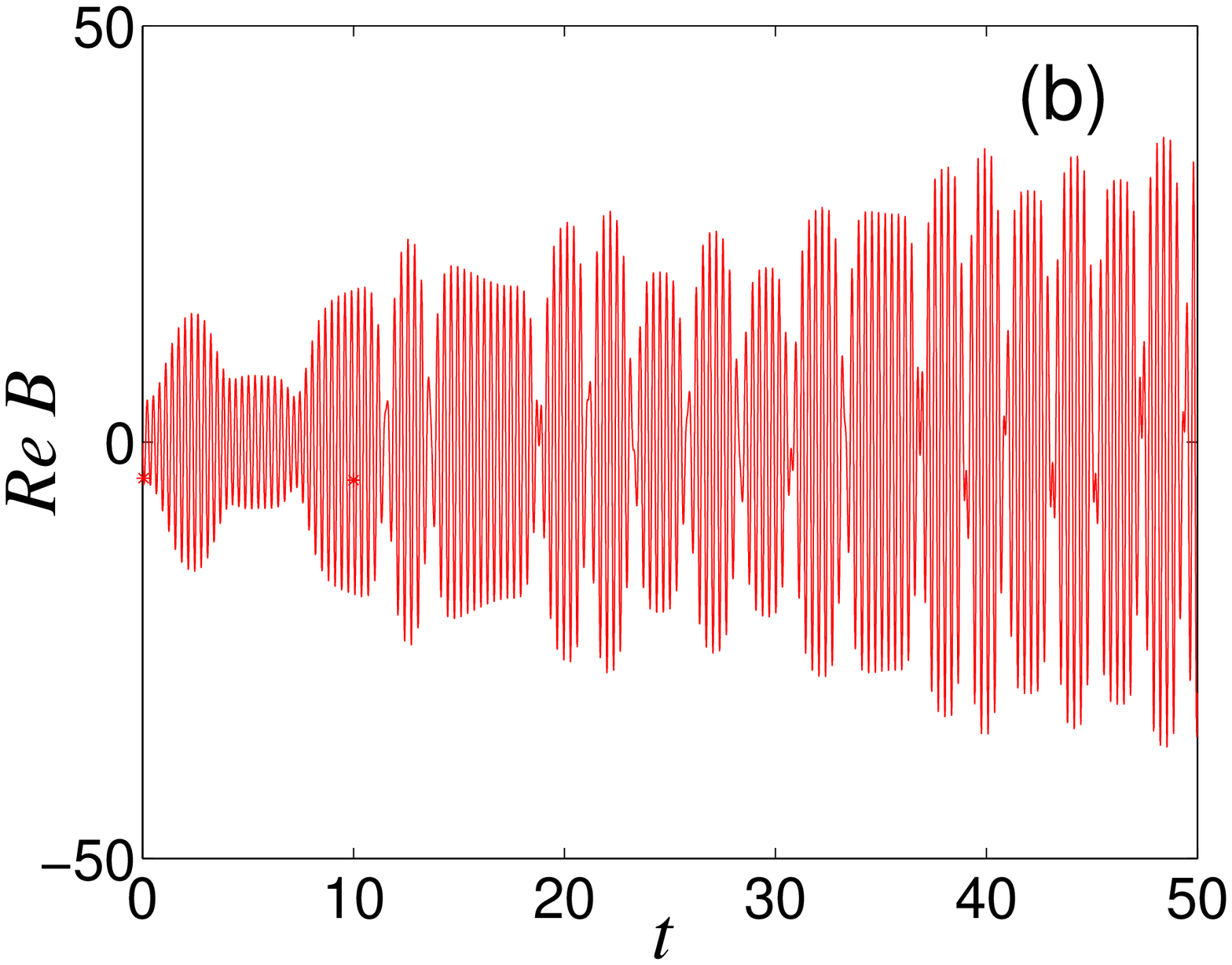}
\includegraphics[angle=90,width=8cm,height=12cm,angle=0]{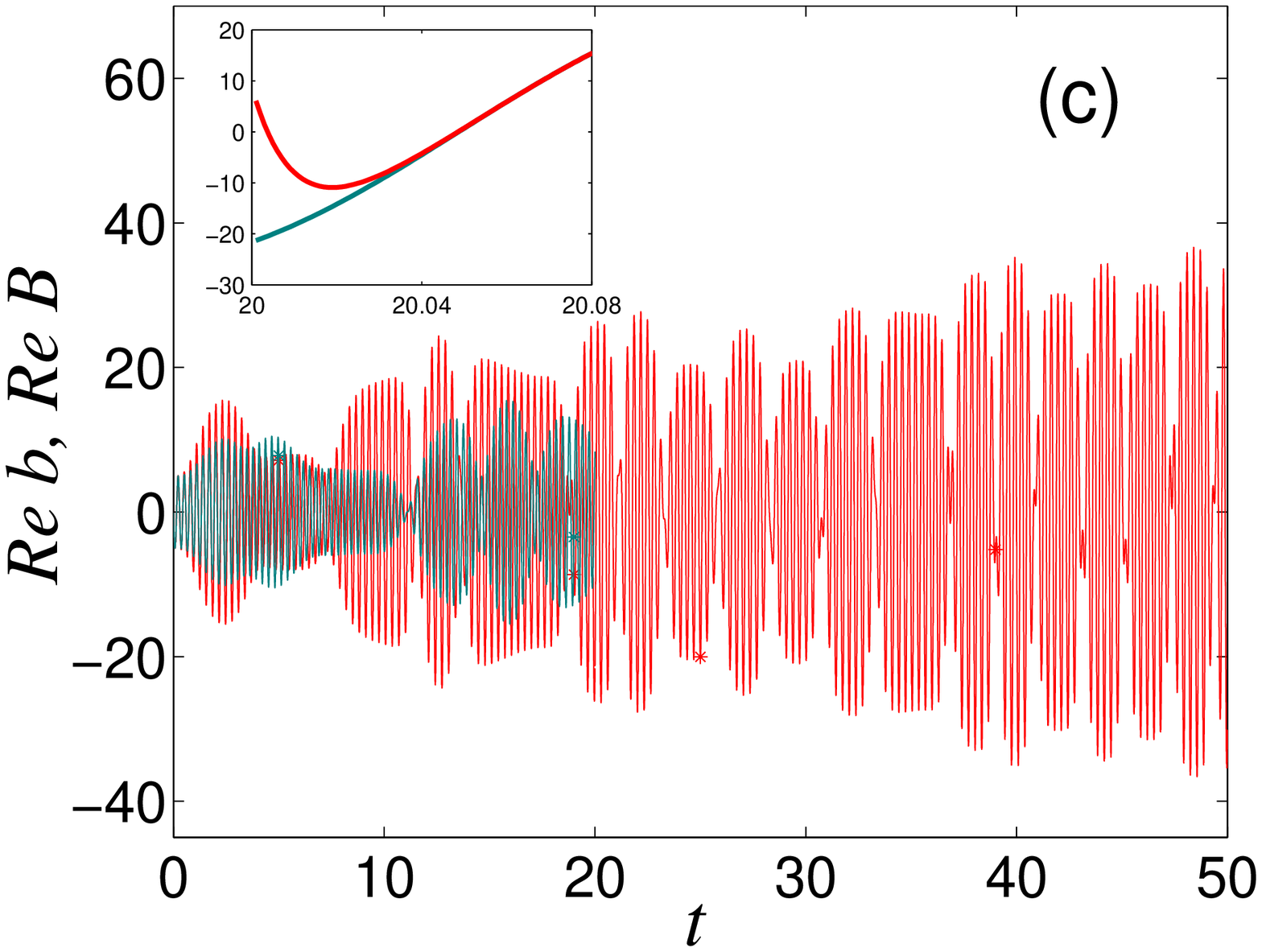}
\includegraphics[angle=90,width=8cm,height=12cm,angle=0]{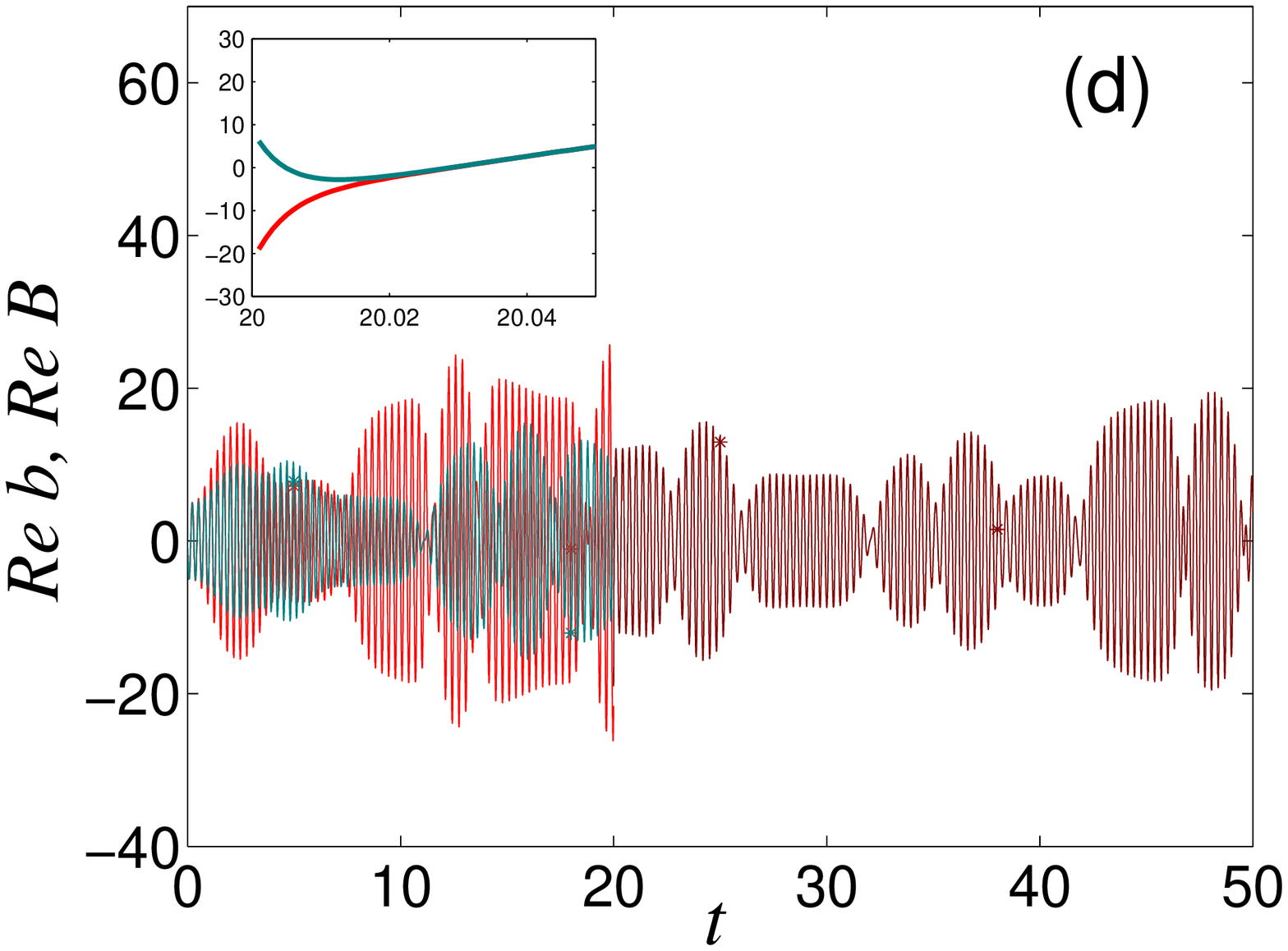}
\caption{}
 \label{fig.3}
\end{figure}

Chaotic beats generated by two independent systems
(\ref{e1})-(\ref{e2}) with different
 values of the amplitudes $F$ and  slightly detuned in frequencies
$\Omega$ and $\Omega+\delta$ can be easily synchronized. As an
numerical example, we consider the following system of four
differential equations:
\begin{eqnarray}
\label{s1}
\frac{da}{dt}&=&-i10a -0.04a +0.1a^{*}b +5e^{-i10.2t}-S_{(a,A)}(a-A)
\,,\\
\label{s2}
\frac{db}{dt}&=&-i20 b  -0.05a^{2}-S_{(b,B)}(b-B)\,,\\
\label{s3}
\frac{dA}{dt}&=&-i10A -0.04A +0.1 A^{*}B +10e^{-i9.9t}-S_{(A,a)}(A-a)\,,\\
\label{s4}
\frac{dB}{dt}&=&-i20 B  -0.05 A^{2}-S_{(B,b)}(B-b)\,.
\end{eqnarray}
As seen, the systems $(a,b)$ and $(A,B)$ are coupled linearly
to each other by the appropriate $S$-terms. The coupling is considered as
 weak if $\omega< S$ (here, $\omega=10$) or as strong if $\omega\geq S$.
 If the coupling can be turned off, that is when $S_{(a,A)}=S_{(b,B)}=S_{(A,a)}=S_{(B,b)}=0$,
  the systems $(a,b)$ and $(A,B)$
 generate  independent chaotic beats (Figs.\ref{fig.3}a and \ref{fig.3}b) characterized
  by appropriate maximal Lyapunov exponents
 $\lambda_{(a,b)}=0.008$ and $\lambda_{(A,B)}=0.007$.
The coupling is
 turned on at an arbitrarily requested  time $t_{on}$.
 Figure \ref{fig.3}c presents the case of unidirectional synchronization [Pyragas, 1992; Pikovsky {\em et al.}, 2001],
 (the coefficients $S_{(A,a)}$ and
 $S_{(B,b)}$ in (\ref{s3})-(\ref{s4}) are equal to zero  whereas the
  terms governed by the coefficients
$S_{(a,A)}$ and
 $S_{(b,B)}$  are turned on at the time $t_{on}=20$. The chaotic beats in
  $(a,b)$-system (receiver)
synchronize with the beats generated by the $(A,B)$-system
(transmiter) -- green vibrations behave exactly as red ones. The
synchronization time is equal  to $t_{s}=0.05$. Let us emphasize that the
 synchronization process is possible if $S_{(A,a)}=S_{(B.b)}\geq 100$
 (which means that synchronization occurs only if the transmiter's  effect
 on the receiver is strong).

The case of mutual synchronization , where $S_{(A,a)}=S_{(B.b)}=S_{(a,A)}=S_{(b.B)}=100$
 is presented in Fig.\ref{fig.3}d. Here, initially different chaotic beats in the systems
 $(a,b)$ and $(A,B)$ uniform their structure -- green and read beats form
  new identical vibrations -- black.
 The synchronization time equals $t_{s}=0.03$ and is nearly twice shorter
  than in the unidirectional synchronization).\\
In conclusion, simple optical systems have been shown to be a possible source of chaotic beats.
\newpage
\centerline{\bf Figure Captions}

Figure 1. Evolution of $Re\, b(t)$
versus $t$ for the system (\ref{e1})-(\ref{e2}).
 The parameters $\omega$, $\epsilon$ and $F$ are constant in
 time: $\omega=10$, $\epsilon=0.1$ and $F=5$. The parameters $\gamma$ and $\Omega$ change
 their values in time:\\
 1) $\Omega=10.5$, $\gamma=0.5$ if $0\leq t< 30$;\\
  2) $\Omega=9.9$, $\gamma=0.5$ if $30\leq t< 80$;\\
  3) $\Omega=9.9$, $\gamma=0.04$ if $80\leq t< 180$;\\
 4) $\Omega=9.9$, $\gamma=0.5$ if $180\leq t<230$;\\
 5) $\Omega=10.5$, $\gamma=0.5$ if $t\geq 230$.\.\\
 The system starts from the initial conditions $a(0)=10$ and $b(0)=-5i$. Enlargements
 $W_{1}$, $W_{2}$  and $W_{3}$ show periodic oscillations, quasiperiodic beats and chaotic beats,
  respectively.\\ \\

Figure 2. The values of maximal Lyapunov exponents marked by an appropriate color for the system (\ref{e1})-(\ref{e2})
with $\omega=10$, $\epsilon=0.1$, $0<\gamma<2 $ and $9<\Omega<11$ \\ \\

Figure 3.(a)-(b):Chaotic beats in the variables $Re\,b(t)$ and
$Re\,B(t)$ of the system (\ref{s1})--(\ref{s4}) if
$S_{(a,A)}=S_{(b,B)}=S_{(A,a)}=S_{(B,b)}=0$. The system starts
from the initial conditions $a(0)=10$, $b(0)=5i$, $A(0)=10$ and
$B(0)=-5i$. (c) Unidirectional synchronization
($S_{(A,a)},S_{(B,b)} $-terms turned off,
$S_{(a,A)},S_{(b,B)}$-terms turned on at the time $t_{on}=20$).
Green beats begin to behave identically to  red ones at the time
$t=0.05$. The window shows exactly the
 beginning of synchronization. (d) Mutual synchronization ( all the $S$-terms turned
 on at $t_{on}=20$). Green and read  beats uniform
 their structure after $t=0.03$ (see, enlargement) turn into black.



\thispagestyle{empty}
\newpage
\thispagestyle{empty}
\newpage
\thispagestyle{empty}
\newpage
\end{document}